\documentclass[aps,prd,twocolumn,groupedaddress,showpacs]{revtex4}
\usepackage{amsmath,amssymb}\begin{document}\title{Geodesic behavior
of sudden future singularities} \author{L. Fern\'andez-Jambrina}
\email[]{lfernandez@etsin.upm.es}
\homepage[]{http://debin.etsin.upm.es/ilfj.htm}
\affiliation{Matem\'atica Aplicada, E.T.S.I. Navales, Universidad
Polit\'ecnica de Madrid,\\
Arco de la Victoria s/n, \\ E-28040 Madrid, Spain}%
\author{Ruth Lazkoz}\email[]{wtplasar@lg.ehu.es} \affiliation{F\'\i
sica Te\'orica, Facultad de Ciencia y Tecnolog\'\i a, Universidad del
Pa\'\i s Vasco,\\ Apdo.  644, E-48080 Bilbao, Spain}
      
\begin{abstract}In this paper we  analyze the effect of recently proposed classes of sudden future
singularities on causal geodesics of FLRW spacetimes.
Geodesics are shown to be extendible and just the equations for
geodesic deviation are singular, although tidal forces are not
strong enough to produce a Big Rip. For the sake of completeness, we compare with the typical sudden future
singularities  of phantom cosmologies.\end{abstract}  

\maketitle
Recently it has been suggested \cite{barrow} that in an expanding FLRW
universe a curvature singularity may appear at a finite time before
Big Crunch for matter contents that satisfy both weak and strong
energy conditions.  This family of models has been further enlarged
\cite{barrow1}, and the same sort of behavior has also been found in
inhomogeneous models \cite{dabrowski}.  It has been remarked, however, that the
dominant energy condition must be violated in order to produce such
sudden singularities \cite{lake}, and that the inclusion of quantum
corrections may appease their strength \cite{odintsov}.  

In these models, the energy density of the formal perfect fluid is
finite at the singularity, but the pressure is infinite.  More
specifically, in the models proposed in \cite{barrow,barrow1} the
scale factor and its first derivative are also finite, whereas second
and higher order derivatives become infinite (in the models presented
in \cite{odintsov} the singularity does not appear in the scale factor
and its first three derivatives are finite).

These sorts of sudden future singularities are quite different from
those in phantom cosmologies \cite{caldwell}, because for the latter
not only does the second derivative of the scale factor blow up at the
singularity, but also do the energy density, the scale factor and its
derivatives from the first order up.

In this paper we want to analyze the behavior of the sudden future
singularities in \cite{barrow,barrow1,dabrowski,lake,odintsov} from a different point of view.  Instead of regarding
the curvature scalar polynomials we shall take a look at causal
geodesics, since they describe the trajectories and the fate of
nonaccelerated observers on these universes.  This is not a difficult
task since FLRW cosmologies, 
{\setlength{\arraycolsep}{1pt} \begin{eqnarray}
&&ds^2=-dt^2+a(t)\left\{f^2(r)dr^2+ r^2\left(d\theta^2+\sin^2\theta
d\phi^2\right)\right\},\nonumber\\
&&f^2(r)=\frac{1}{1-kr^2},\quad k=0,\pm1,\end{eqnarray}} 
\noindent are homogeneous
and isotropic and therefore have a six-dimensional group of
isometries generated, for instance, by the Killing
fields
{\setlength{\arraycolsep}{1pt}\begin{subequations}
\begin{eqnarray}\xi_{1}&=&\frac {\sin
\theta \cos \phi }{f(r) }\,\partial_{r} +\frac {\cos\theta \cos \phi
}{rf(r)}\,\partial_{\theta}-\frac {\sin \phi}{rf ( r) \sin \theta
}\,\partial_{\phi},~~~\\\xi_{2}&=&\frac {\sin \theta \sin\phi }{f(r)
}\,\partial_{r} +\frac {\cos\theta \sin \phi
}{rf(r)}\,\partial_{\theta}+\frac {\cos \phi}{rf ( r) \sin \theta
}\,\partial_{\phi},~~~\\\xi_{3}&=&\frac{\cos\theta}{f(r)}\,\partial_{r}-\frac{\sin\theta}{rf(r)}\,\partial_{\theta},~~~\\\zeta_{1}&=&\cos\phi\,\partial_{\theta}-\cot\theta
\sin\phi\,\partial_{\phi},~~~\\\zeta_{2}&=&\sin\phi\,\partial_{\theta}+\cot\theta
\cos\phi\,\partial_{\phi},~~~\\\zeta_{3}&=&\,\partial_{\phi},~~~\end{eqnarray}
\end{subequations}}
which yield six different constants of geodesic motion, i.e. three linear momenta
and three angular momenta:
\begin{subequations}\begin{eqnarray}
P_{1}&=&a(t)\left\{\frac {
r}{f(r)}\left(\cos \theta\cos \phi \,\dot\theta-\sin \theta \sin \phi
\,\dot \phi\right) +\right.\\\nonumber&&\left.
f(r) \sin \theta \cos \phi \,\dot r\right\},\\
P_{2}&=&a(t) \left\{\frac {
r}{f(r)}\left(\cos \theta\sin \phi\,\dot\theta+\sin \theta \cos \phi
\,\dot \phi\right)+\right.\\&&\left.f(r) \sin \theta \sin \phi \,\dot r\right\},\\
P_{3}&=&a(t)\left( f(r) \cos \theta \,\dot r-\frac { r }{f(r)}\sin
\theta\,\dot\theta\right),\\L_{1}&=&a(t) r^2\left(\cos \phi
\,\dot\theta-\sin \theta \cos\theta\sin \phi \,\dot\phi\right),\\
L_{2}&=&a(t) r^2\left(\sin \phi \,\dot\theta+ \sin \theta
\cos\theta\cos \phi \,\dot\phi\right),\\L_{3}&=&a(t)r^2 \sin^2
\theta\,\dot\phi,
\end{eqnarray}
\end{subequations}
for a geodesic parametrized by its proper time $\tau$, so that
$d\tau^2=-ds^2$.  The dots stand for derivation with respect to this
proper time.  We define now
\begin{equation}\delta\equiv\dot
t^2-a(t)\left\{f^2(r)\,\dot r^2+ r^2\left(\dot\theta^2+\sin^2\theta\,
\dot\phi^2\right)\right\},
\end{equation}
where $\delta$ is zero for null geodesics and one for timelike
geodesics.  With such an amount of conserved quantities, geodesic
equations reduce to first order differential
equations:
\begin{subequations}
\begin{eqnarray}\dot t^2&=&\delta +
\frac{P^2+k L^2}{a(t)},\\\dot r&=&\frac {P_{1}\,\sin \theta \cos \phi
+P_{2}\,\sin \theta \sin \phi +P_{3}\,\cos\theta}{a(t)f(r)},\\\dot
\theta&=&\frac {L_{1}\cos\phi+L_{2}\sin \phi
}{a(t)r^2},\\\dot\phi&=&\frac {L_{3}}{a(t)r^2 \sin^2
\theta},
\end{eqnarray}
\end{subequations}
in terms of total linear momentum and angular momentum
\begin{equation}
P^2=P_{1}^2+P_{2}^2+P_{3}^2,\qquad
L^2=L_{1}^2+L_{2}^2+L_{3}^2.
\end{equation} 
The system may be further
simplified, since due to spherical symmetry every geodesic may be fit
in the hypersurface $\theta=\pi/2$, with $L_{1}=L_{2}=0=P_{3}$, by a
suitable choice of the coordinates,
then
\begin{subequations}
\begin{eqnarray}\dot t^2&=&\delta +
\frac{P^2+k L^2}{a(t)},\\\dot r&=&\frac {P_{1}\,\cos \phi +P_{2}\,
\sin \phi}{a(t)f(r)},\\\dot\phi&=&\frac
{L_{3}}{a(t)r^2},
\end{eqnarray}
\end{subequations}
It can be easily noticed that these equations are singular if and only if
$a(t)$ has a zero, which corresponds to either a Big Bang or a Big
Crunch singularity.  Therefore, if we consider models with sudden
future singularities like those in
\cite{barrow},
\begin{equation}a(t)=1+
\left(\frac{t}{t_{s}}\right)^q(a_{s}-1)-\left(1-\frac{t}{t_{s}}\right)^n,\label{abarrow}
\end{equation}
with constants $a_{s}$, $t_{s}$, $0<q\le 1$, $1<n<2$, we realize that
the geodesics just see the Big Bang singularity at $t=0$, but not the
sudden singularity at $t=t_{s}$, where the scale factor does not
vanish.  This is obvious, since these universes are
$C^1$-differentiable manifolds but for the Big Bang.  

Generalizations to (\ref{abarrow}) have been also considered. For instance, in \cite{barrow1},  the following evolution was put forward (among others):
\begin{equation}a(t)=a_{s}-1+{\rm exp}{(\lambda(t-t_s))}-\left(1-\frac{t}{t_{s}}\right)^n,
\end{equation}
with $\lambda>0$ and $n$ in the same range as above. 
Similarly, in \cite{odintsov}  a quantum inspired model was proposed for which $a(t)$  has functionally the form of (\ref{abarrow}), but with \mbox{$3<n<4$} instead, so that these universes are
$C^3$-differentiable manifolds but for the Big Bang.

Furthermore,
since in these settings $a$, $a'$ are finite at $t_{s}$ and the singularity appears just in higher order derivatives of $a$, the
acceleration vector of the geodesic, $(\ddot t, \ddot r, \ddot\theta,
\ddot \phi)$, which comprises the effect of inertial forces, is also
regular.  Only the third derivative of the parametrization of the
geodesic is singular at $t_{s}$, but we just require first and second
derivatives to define geodesic equations.  Causal geodesics in such
universes do not see the singularities but through geodesic deviation
effects, since they are due to the Riemann tensor.  Point particles
travelling along causal geodesics do not experience any singularity,
but extended objects might suffer infinite tidal forces at $t=t_{s}$.

According to Tipler's definition \cite{tipler} a strong curvature
singularity is encountered at a point $p$ if every volume element defined by
three linearly independent, vorticity-free, geodesic deviation vectors along
every causal geodesic through $p$ vanishes at this point. This
definition comes to say that an extended finite object is
crushed to zero volume by tidal forces at a strong singularity.
Generalizations of this widely accepted definition may be found in
\cite{krolak, rudnicki}.

In \cite{clarke}, necessary and sufficient conditions for the
appearance of strong curvature singularities are shown. For instance, 
if a causal geodesic meets a strong singularity at a value $\tau_{s}$ 
of its affine parametrization, expressions of the form
\begin{equation}\int_{0}^{\tau}d\tau'\int_{0}^{\tau'}d\tau''|R^{i}_{\ 
    0j0}(\tau'')|\;,\end{equation}
will diverge along the geodesic on approaching $\tau_{s}$. The components of the Riemann tensor are
understood to be written in a frame parallely transported along the
geodesic. Similar results involving double integrals of the component 
$R_{00}$ of the Ricci tensor or triple integrals of components
$C^{i}_{\ 0j0}$ of the Weyl tensor are written for lightlike geodesics.

For Krolak's definition, necessary conditions are milder, since they
involve a simple integral of components of the curvature tensor:
\begin{equation}\int_{0}^{\tau}d\tau'|R^{i}_{\ 
    0j0}(\tau')|\;.\end{equation}
For null geodesics conditions are relaxed in a similar way.

In the case of  the sudden singularities in \cite{barrow,barrow1} the components of the Riemann
tensor diverge as $a''$, since $a'$ and $a$ are finite;  and in the
worst case they diverge as a power $n-2$, for $1<n<2$. Therefore after
one integration of the components of the Riemann tensor, the power
will be positive and the integral will not diverge. Of course, the
situation is even more favorable if singularities do not arise in $a''$
but in higher derivatives like in those in \cite{odintsov}.

Hence we have shown that sudden singularities are not strong according
to Tipler and Krolak's definitions and therefore tidal forces do not
crush all finite bodies. This is quite important, since it means that 
the spacetime may be extended across sudden singularities
\cite{tipler} and cannot be considered the final fate of these
universes.

Let us come to conclusions now. In this paper we have shown that causal geodesics are not affected by the sudden future singularities in some recently put forward models, since these singularities are not seen by geodesic
equations. Recall that geodesic incompleteness is the standard definition for
singularities in General Relativity \cite{HE}. 

Furthermore, considering just curvature singularities, it has been
shown that they are weak according to Tipler's and Krolak's definitions,
and therefore finite objects are not necessarily torn on crossing the 
singularities.

In contrast, since in the typical sudden future singularities of phantom cosmologies there is a blow up of the scale factor and all its derivatives, such singularities are indeed seen by geodesic equations, thus altering causal geodesics, and leading to destruction of structure (or Big Rip) \cite{caldwell,nes}. 

\section*{Acknowledgments}R.L. is supported by  the University of the Basque Country through research grant 
UPV00172.310-14456/2002 and by the Spanish Ministry of Education and Culture  through research grant  FIS2004-01626.

\end{document}